\documentclass[twoside]{ilcws08}
\usepackage[latin1]{inputenc}
\usepackage[dvips]{graphicx,epsfig,color}
\usepackage{wrapfig,rotating}
\usepackage{amssymb,amsmath,array}

\begin{document}
\title{The ILC DEPFET Prototype: Report of the Test Beam at CERN 2008} %% 

%***********************************************************************
% AUTHORS INFORMATION AREA
%***********************************************************************
\author{C. Mari\~nas
% Optional short acknowledgment: remove next line if non-needed
\thanks{On behalf of the DEPFET collaboration.}
% DO NOT MODIFY THE FOLLOWING '\vspace' ARGUMENT
\vspace{.3cm}\\
% Addresses and institutions (remove "1- " in case of a single institution)
Instituto de F\'isica Corpuscular (IFIC), CSIC-UVEG\\
%Dept. F\'isica At., Mol. y Nuclear, Univ. of Valencia\\
P.O. Box 22085. Valencia - Spain
%% Remove the next three lines in case of a single institution
}
%%***********************************************************************
% END OF AUTHORS INFORMATION AREA
%***********************************************************************

\maketitle

\hyphenation{de-mons-tra-ted}
\hyphenation{subs-tra-te}
\hyphenation{co-llec-ted}
\hyphenation{pro-ba-ble}

\begin{abstract}
The DEPFET Collaboration pursues the development of a high resolution pixel vertex detector for future colliders (like ILC), based on the integration of amplifying transistors into a fully depleted bulk. In August 2008, six DEPFET prototypes were tested in a pion beam at SPS complex at CERN, collecting more than 20 million of events. In this contribution, the prototype system, the experimental setup, the analysis software and preliminary results are presented. 

\end{abstract}

\section{Introduction}

DEPFET (DEpleted P-channel Field Effect Transistor) based detectors are currently developed for X-ray astronomy~\cite{XEUS}, biomedical autoradiography~\cite{medical} as well as for vertex detectors in future $e^{+}$$e^{-}$ colliders.
The latter application requires excellent vertex reconstruction and efficient heavy quark flavour tagging using low momentum tracks. These requirements impose unprecedented constraints on the detector: High granularity, fast read-out, low material budget and low power consumption. Measurements on realistic DEPFET prototypes have demonstrated that the concept is one of the principal candidates~\cite{options,performance} to meet these challenging requirements.

\section{DEPFET principle and operation}

In the DEPFET active pixel detectors~\cite{operation}, each pixel consists of a p-channel Field Effect Transistor integrated in a fully depleted bulk. Free charge carriers created in the substrate by ionizing particles drift towards a deep implant underneath the transistor channel. The charge trapped in the internal gate modulates the transistor current. Thus, the amplification of the signal is achieved in the sensor. After read-out the accumulated signal is removed from the internal gate by applying a positive voltage on a clear contact. A matrix of DEPFET pixels is read out using the {\em rolling shutter} concept~\cite{system}.

\section{The ILC prototype system}

The ILC DEPFET prototype system consists of two mayor parts: The hybrid and the r/o boards. The hybrid board hosts the DEPFET matrix (128x64 pixels, from the most recent production PXD5), two steering chips (the so-called Clear/Gate Switchers that address the rows for read-out and clear) and the readout chip (CUrrent ReadOut, with on-chip pedestal substraction). The readout board takes care of the configuration of all the chips, the digitization and storage of the analog data as well as the communication with the PC (wich runs the DAQ program) via an USB board. This prototype system is analog to the one used in past Test Beams~\cite{TB1}~\cite{TB2}.

\begin{wrapfigure}{r}{0.5\columnwidth}
\vspace{-12pt}
\centerline{\includegraphics[width=0.5\columnwidth]{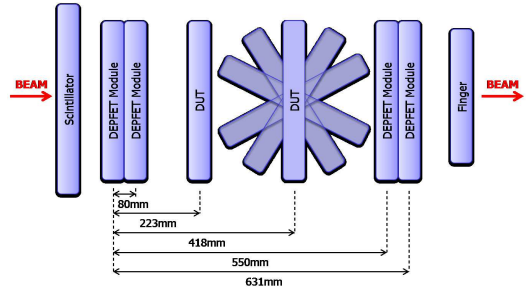}}
\vspace{-5pt}
\caption{The set-up of the DEPFET telescope in the CERN SPS H6B area}\label{Fig:set_up}
\end{wrapfigure}

\section{Test Beam at CERN}

In August 2008 beam test measurements were carried out at the SPS complex at CERN, using 120 GeV/c pions. All the modules in the beam test have been fully characterized using laser and radioactive sources and the electrical settings optimized for the best performance. The set-up is shown in the Figure \ref{Fig:set_up}. 5 DEPFET planes (matrices with 128x64 pixels of size 32x24$\mu$$m^{2}$ and 450$\mu$m thick) were used as telescope, to reconstruct the tracks and predict the impact position in the Device Under Test (DUT). In the middle of the telescope, the DUT (128x64 pixels, size 24x24$\mu$$m^{2}$, 450$\mu$m thick and working fully depleted) was placed in a rotating motorstage. The synchronization of all the system was made by a TLU (Trigger Logit Unit)~\cite{TLU}.

\section{Preliminary results}

The data from the 2008 test beam were analyzed using the software framework developed by EUDET~\cite{EUTelescope}. The DEPFET raw data were converted from a proprietary format to the standard Linear Collider I/O (LCIO) format, pedestal and common mode corrections were applied, the signal on adjacent pixels was clustered. The position of clusters was determined using the center-of-gravity method with an $\eta$-correction. The alignment of the telescope modules was determined using the Millipede algorithm~\cite{millepede}. Finally, the position of the particle on the DUT is predicted using a track fit to the telescope hits.

\begin{wrapfigure}{r}{0.5\columnwidth}
%\centerline{\includegraphics[viewport= 20 100 335 450,angle=90,width=0.38\columnwidth,height=0.25\textheight]{sig3x3_run1258_recortada.eps}}
\vspace{-16pt}
\centerline{\includegraphics[width=0.5\columnwidth,height=0.25\textheight]{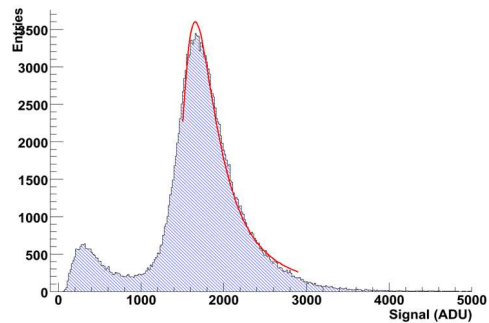}}
%\centerline{\includegraphics[viewport= -20 -50 180 180,width=0.3\columnwidth,height=0.1\textheight]{sig3x3_run1258_recortada.eps}}
%\vspace{10pt}
\caption{Signal collected by a 3x3 cluster for normal incidence of 120 GeV/c pions on a DEPFET DUT in nominal conditions. 7$\sigma$ and 3$\sigma$ cuts for seed and neighbours.}\label{Fig:signal}
\vspace{-10pt}
\end{wrapfigure}

The result obtained for the signal collected in a 3x3 cluster for normal beam incidence is shown in Figure \ref{Fig:signal}. The most probable deposited charge of a MIP is $\thicksim$1700 ADC units. Combining this value with an average noise of $\thicksim$12.5 ADC units, the signal over noise ratio (SNR) is 135 for a DUT of 450$\mu$m ($g_{q}$$\approx$360pA/$e^{-}$).

\begin{wrapfigure}{r}{0.5\columnwidth}
%\vspace{-70pt}
%\centerline{\includegraphics[width=0.5\columnwidth, height=0.4\textheight]{resolY.eps}}
\centerline{\includegraphics[width=0.5\columnwidth, height=0.3\textheight]{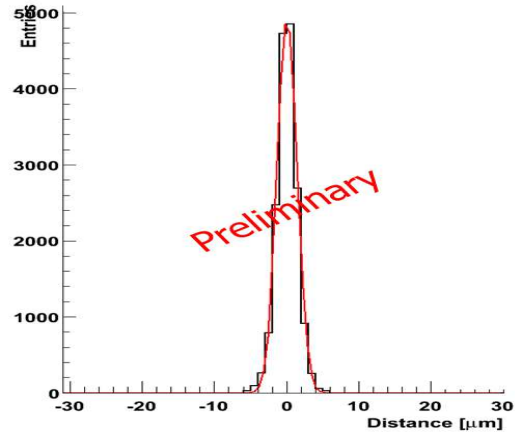}}
\vspace{-10pt}
\caption{$\sigma^{2}_{Total}=\sigma^{2}_{Telescope}+\sigma^{2}_{DUT}+\sigma^{2}_{m.s.}$. Total resolution in Y direction for DUT.}\label{Fig:residual}
\vspace{-100pt}
\end{wrapfigure}

A typical residual distribution (the difference between predicted position and the DUT measurement) is shown in Figure \ref{Fig:residual}. The width of this distribution is a measure of the intrinsic DUT resolution, but also receives sizeable contributions due to multiple scattering and the resolution of the telescope. The width of the residual distribution was measured to be 1.94 in the y-direction and 2.84 mm in X. The difference between the results in both coordinates is due to the rectangular pixels in the telescope.

%\vspace{3cm}

\section{Conclusions}

A telescope made up of six DEPFET planes from the PXD5 sensor production was operated successfuly in the SPS beam line in August 2008. On the 450$\mu$m thick device under test with 24x24$\mu$$m^{2}$ pixels the most probable signal was found to be 17000 ADC units, compared to a pixel noise of 12.5 ADC counts. The intrinsic resolution of this device was found to be better than 1.94$\mu$m. The analysis is in progress and final results will be presented in a future.

\vspace{2cm}

\section{Bibliography}

% ****************************************************************************
% BIBLIOGRAPHY AREA
% ****************************************************************************

\begin{footnotesize}

\end{footnotesize}

% ****************************************************************************
% END OF BIBLIOGRAPHY AREA
% ****************************************************************************

\end{document}